\documentclass[sigconf]{aamas}

\usepackage{tabularx}
\usepackage{graphicx}
\usepackage{soul}
\usepackage{url}
\usepackage{hyperref} 
\usepackage[utf8]{inputenc}
\usepackage{titlesec}
\usepackage{booktabs}

\usepackage{makecell} 
\usepackage{mdframed}

\usepackage{enumitem}
\usepackage{graphicx}
\usepackage{wrapfig}
\usepackage{array}
\usepackage{balance} 
\usepackage{todonotes}
\usepackage{tikz}
\usetikzlibrary{arrows.meta, positioning, shapes, shadows}
\usepackage{makecell}
\usepackage{graphicx}
\usepackage{tikz}
\usetikzlibrary{arrows.meta,positioning}

\makeatletter
\gdef\@copyrightpermission{
  \begin{minipage}{0.2\columnwidth}
   \href{https://creativecommons.org/licenses/by/4.0/}{\includegraphics[width=0.90\textwidth]{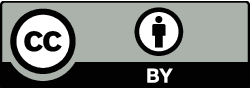}}
  \end{minipage}\hfill
  \begin{minipage}{0.8\columnwidth}
   \href{https://creativecommons.org/licenses/by/4.0/}{This work is licensed under a Creative Commons Attribution International 4.0 License.}
  \end{minipage}
  \vspace{5pt}
}
\makeatother

\setcopyright{ifaamas}
\acmConference[AAMAS '26]{Proc.\@ of the 25th International Conference
on Autonomous Agents and Multiagent Systems (AAMAS 2026)}{May 25 -- 29, 2026}
{Paphos, Cyprus}{C.~Amato, L.~Dennis, V.~Mascardi, J.~Thangarajah (eds.)}
\copyrightyear{2026}
\acmYear{2026}
\acmDOI{}
\acmPrice{}
\acmISBN{}

\title[AAMAS-2026 Formatting Instructions]{Strategic Interactions in Multi-Level Stackelberg Games with Non-Follower Agents and Heterogeneous Leaders}
\subtitle{Extended Abstract}

\author{Niloofar Aminikalibar}
\affiliation{
  \institution{Aston University}
  \city{Birmingham}
  \country{United Kingdom}}
\email{namin21@aston.ac.uk}

\author{Farzaneh Farhadi}
\affiliation{
  \institution{Aston University}
  \city{Birmingham}
  \country{United Kingdom}}
\email{f.farhadi@aston.ac.uk}

\author{Maria Chli}
\affiliation{
  \institution{Aston University}
  \city{Birmingham}
  \country{United Kingdom}}
\email{m.chli@aston.ac.uk}

\begin{abstract}
Strategic interaction in congested systems is commonly modelled using Stackelberg games, where competing leaders anticipate the behaviour of self-interested followers. A key limitation of existing models is that they typically ignore agents who do not directly participate in market competition, yet both contribute to and adapt to congestion. Although such non-follower agents do not generate revenue or respond to market incentives, their behaviour reshapes congestion patterns, which in turn affects the decisions of leaders and followers through shared resources.

We argue that overlooking non-followers leads to systematically distorted equilibrium predictions in congestion-coupled markets. To address this, we introduce a three-level Stackelberg framework with heterogeneous leaders differing in decision horizons and feasible actions, strategic followers, and non-follower agents that captures bidirectional coupling between infrastructure decisions, competition, and equilibrium congestion.

We instantiate the framework in the context of electric vehicle (EV) charging infrastructure, where charging providers compete with rivals, while EV and non-EV traffic jointly shape congestion. The model illustrates how explicitly accounting for non-followers and heterogeneous competitors qualitatively alters strategic incentives and equilibrium outcomes. Beyond EV charging, the framework applies to a broad class of congestion-coupled multi-agent systems in mobility, energy, and computing markets.

\end{abstract}


\newcommand{\BibTeX}{\rm B\kern-.05em{\sc i\kern-.025em b}\kern-.08em\TeX}

\begin{document}


\pagestyle{fancy}
\fancyhead{}


\maketitle 

\section{Introduction}

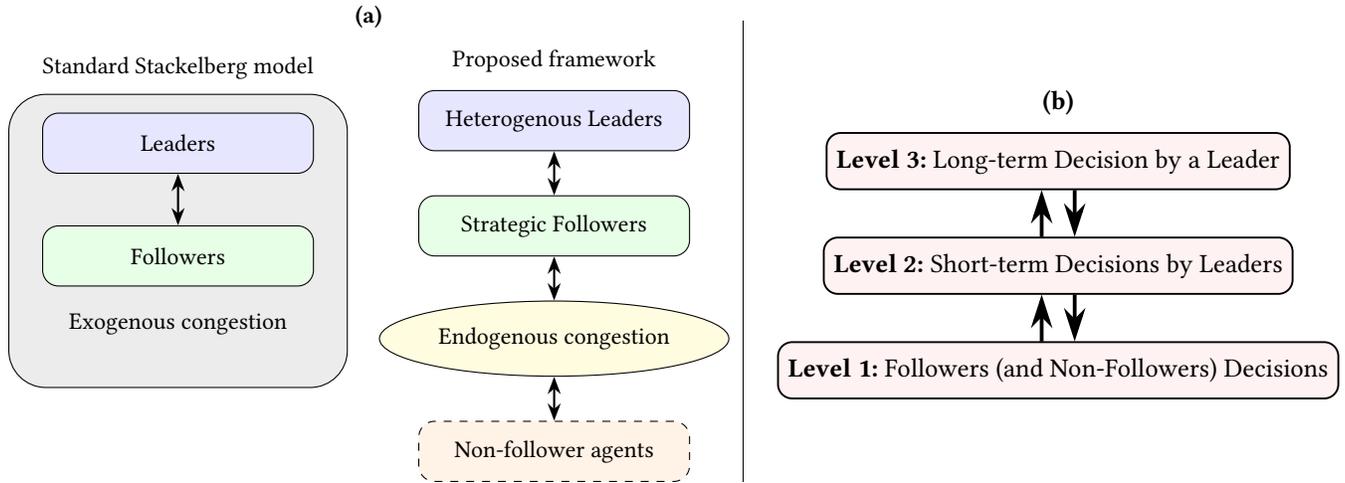
\begin{figure*}[t]
\centering
\begin{minipage}{0.55\textwidth}
    \centering
\begin{tikzpicture}[
    box/.style={
        draw,
        rectangle,
        rounded corners=6pt,
        minimum width=3.6cm,
        minimum height=0.8cm,
        align=center,
        font=\normalsize
    },
    dashedbox/.style={
        draw,
        rectangle,
        dashed,
        rounded corners=6pt,
        minimum width=3.6cm,
        minimum height=0.8cm,
        align=center,
        font=\normalsize
    },
    oval/.style={
        draw,
        ellipse,
        minimum width=4.4cm,
        minimum height=1 cm,
        align=center,
        font=\normalsize
    },
    arrow/.style={<->, thick, >=Stealth}
]


\node[
    draw,
    fill=gray!15,
    rounded corners=14pt,
    minimum width=4.5cm,
    minimum height=3.9cm
] (bg) at (-3.2,0.7) {};

\node[box, fill=blue!10]  (L1) at (-3.2,2.0) {Leaders};
\node[box, fill=green!10] (F1) at (-3.2,0.5) {Followers};

\draw[arrow] (L1.south) -- (F1.north);

\node[font=\normalsize] at (-3.2,-0.4) {Exogenous congestion};
\node[font=\normalsize] at (-3.2,3) {Standard Stackelberg model};
\node[font=\normalsize] at (1.8,3.1) {Proposed framework};

\node[box, fill=blue!10]   (L2) at (1.8,2.3) {Heterogenous Leaders};
\node[box, fill=green!10]  (F2) at (1.8,0.9) {Strategic Followers};
\node[oval, fill=yellow!15] (C2) at (1.8,-0.6) {Endogenous congestion};
\node[dashedbox, fill=orange!10] (N2) at (1.8,-2.1) {Non-follower agents};

\draw[arrow] (L2.south) -- (F2.north);
\draw[arrow] (F2.south) -- (C2.north);
\draw[arrow] (C2.south) -- (N2.north);
\node[anchor=south, font=\bfseries]
  at ([yshift=2pt]current bounding box.north) {(a)};
\end{tikzpicture}
\end{minipage}
 \vrule height 3cm width 0.2pt
\hfill
\begin{minipage}{0.42\textwidth}
\centering
\resizebox{\linewidth}{!}{%
\begin{tikzpicture}[scale=0.7, transform shape,
    node distance=5mm and 4mm,
    box/.style={rectangle, draw=black, rounded corners=3pt, minimum width=0.4\linewidth, minimum height=6mm, align=center, font=\small, fill=pink!20},
    arrdown/.style={-{Stealth[length=2.5mm,width=1.6mm]}, thick},
    arrup/.style={{Stealth[length=2.5mm,width=1.6mm]}-, thick}
]

\node[box] (L1) {\textbf{Level 3:} Long-term Decision by a Leader};
\node[box, below=of L1] (L2) {\textbf{Level 2:} Short-term Decisions by Leaders};
\node[box, below=of L2] (L3) {\textbf{Level 1:} Followers (and Non-Followers) Decisions};

\draw[arrdown] ([xshift=-5pt]L2.north) -- ([xshift=-5pt]L1.south);
\draw[arrup]   ([xshift=5pt]L2.north)  -- ([xshift=5pt]L1.south);

\draw[arrdown] ([xshift=-5pt]L3.north) -- ([xshift=-5pt]L2.south);
\draw[arrup]   ([xshift=5pt]L3.north)  -- ([xshift=5pt]L2.south);
\node[anchor=south, font=\bfseries]
  at ([yshift=2pt]current bounding box.north) {\small(b)};
\end{tikzpicture}%
}
\end{minipage}

\caption{\textbf{(a):} \small Conceptual comparison between a standard and proposed Stackelberg framework with non-follower agents. Existing approaches typically treat non-participating agents and background traffic as exogenous, neglecting their adaptive response to congestion. Our framework explicitly models non-follower agents whose behaviour both contributes to and adapts to congestion, reshaping equilibrium incentives for strategic leaders and followers.  \textbf{(b):} Three-level hierarchical Stackelberg framework for heterogeneous leaders.}
\label{fig:combined_figures}
\end{figure*}

Many real-world multi-agent systems involve strategic competition over resources whose value depends on congestion. Examples include transportation networks, energy systems, cloud computing, and digital marketplaces. In such settings, providers make long-term and short-term strategic decisions while anticipating how large user populations respond to prices, delays, and capacity constraints.
Stackelberg games have become a standard modelling tool for capturing this interaction, representing providers as leaders and users as followers who respond selfishly to announced decisions. 

Despite their widespread use, most Stackelberg models in congested environments rely on a restrictive abstraction: they focus exclusively on agents who directly participate in the leader–follower interaction. Agents who do not generate revenue and do not respond to strategic incentives are typically treated as exogenous background flow or ignored altogether. However, in congestion-coupled systems, these non-follower agents are neither passive nor unaffected: they reshape congestion faced by strategic agents, while simultaneously adjusting their own behaviour to avoid resources made congested by leaders’ and followers’ actions. This bidirectional coupling alters equilibrium structure despite non-followers not participating in market competition.

This paper argues that ignoring non-follower agents is not a benign modelling simplification. When congestion is endogenously determined, background agents can alter equilibrium structure, shift competitive pressure, and change optimal strategic decisions by leaders. As a result, models that abstract away these agents risk producing qualitatively misleading conclusions about profitability, infrastructure deployment, and competitive intensity in markets where congestion is shared across heterogeneous user groups.

We address this gap by introducing a hierarchical Stackelberg framework that integrates three interacting components: heterogeneous leaders operating with distinct decision horizons and action sets; strategic followers who participate in market competition and adjust their decisions under congestion and leadership actions; and non-follower agents who both shape and adapt to congestion without engaging in market competition (see Figure~\ref{fig:combined_figures}(a)). The resulting model captures the bidirectional coupling between long-term decisions, short-term competitive behaviour, and equilibrium congestion within a single conceptual structure.

The framework is demonstrated on electric vehicle (EV) charging infrastructure, a domain of growing importance driven by rapid EV adoption \cite{UKGov2021_EVCharging,UKGov2022_EVStrategy}. In this setting, providers compete for users while charging and travel decisions interact with traffic congestion, and non-EV traffic both generates and adapts to congestion without participating in the charging market. Existing models typically examine either provider competition \cite{RN219,ghavami2023decentralized,zhang2020plug,RN198} or charging–traffic interaction \cite{ge2024distributed,pan2024competitive}, but seldom both, relying on assumptions such as homogeneous providers, fixed locations, or exogenous background traffic. Although EV charging motivates the framework, explicitly modelling non-follower agents and endogenous congestion yields a more faithful account of strategic interaction in congestion-coupled markets.

\section{Problem Description}\label{sec:model}
We study a hierarchical decision problem in which a leader makes a long-term commitment upon entry into a market of competing leaders who interact primarily through short-term decisions. This commitment shapes follower behaviour, influences incumbents’ competitive responses, and affects network congestion. Alongside followers, non-follower agents adapt to congestion without engaging in market competition, yet jointly determine the equilibrium congestion that feeds back into strategic outcomes.

In the EV charging instantiation, a new entrant selects charger locations as a long-term infrastructure decision, while incumbent providers compete through short-term pricing. EV drivers (followers) choose routes and charging locations under congestion, trading off travel time, delay, and price \cite{NiloofarIEEE,niloofarPLMR,niloofarSPRINGER}. Non-EV drivers (non-followers) choose routes to avoid congestion; through shared network usage, EV and non-EV traffic jointly generate and respond to congestion.

Leaders are heterogeneous: incumbents possess fixed infrastructure and compete only on price, whereas the entrant decides on both placement and pricing. The resulting system tightly couples infrastructure deployment, price competition, and congestion-dependent traffic equilibria. The central challenge is to determine how an entrant can deploy infrastructure while anticipating downstream competition and the redistribution of EV and non-EV traffic.

\section{Conceptual Framework}\label{sec:solution}
The interaction described above induces a three-level Stackelberg structure with sequential decision-making, illustrated in Figure~\ref{fig:combined_figures}(b). At the lowest level (Level 1), EV and non-EV drivers respond to prices and congestion by selecting routes and charging options that minimise individual cost. Although non-EV drivers do not participate in the charging market, their routing decisions adapt to congestion generated by charging-related traffic and, in turn, affect travel times and charging choices of EV users. This level captures the bidirectional coupling between different driver populations through shared congestion.

At the intermediate level (Level 2), charging providers compete by setting prices while anticipating how drivers’ congestion-dependent choices respond. Providers’ pricing incentives are shaped by both strategic rivals and non-EV traffic that affects demand through congestion rather than direct interaction.

At the highest level (Level 3), the entrant selects charger locations while anticipating the induced
pricing behaviour of incumbents and the resulting equilibrium traffic patterns. This long-term decision anticipates the induced responses of providers and drivers under congestion.

Together, these layers define a hierarchical Stackelberg structure in which infrastructure decisions, competitive behaviour, and congestion outcomes are jointly determined. The framework treats congestion as an endogenous outcome shaped by strategic decisions and adaptive behaviour across heterogeneous agent populations, rather than as an exogenous background condition.

To operationalise this framework, we adopt a solution approach that aligns with the hierarchical structure of the game. Each decision layer is analysed using equilibrium concepts natural to that layer, while higher-level decisions explicitly anticipate the induced responses downstream. This hierarchical resolution enables principled, equilibrium-consistent reasoning about infrastructure placement and competitive behaviour under congestion, without relying on restrictive assumptions such as agent homogeneity or exogenous traffic.

\section{Acknowledgement}
This paper has been accepted for publication in Proceedings of the 25th International Conference on Autonomous Agents and Multiagent Systems (AAMAS 2026). The final published version is available via the ACM Digital Library.

\bibliographystyle{ACM-Reference-Format} 
\bibliography{sample}

\end{document}